\documentclass[12pt]{elsarticle}

\usepackage{graphicx}
\usepackage{amssymb,amsthm}

\begin{document}

\begin{frontmatter}

\title{Network induces burst synchronisation in cat brain}

\author{Ewandson L. Lameu$^1$, Fernando S. Borges$^1$, Rafael R. Borges$^1$, 
Antonio M. Batista$^{1,2,*}$, Murilo S. Baptista$^3$, Ricardo L. Viana$^4$}
\address{$^1$P\'os-Gradua\c c\~ao em Ci\^encias/F\'isica, Universidade 
Estadual de Ponta Grossa, 84030-900, Ponta Grossa, PR, Brazil.}
\address{$^2$Departamento de Matem\'atica e Estat\'istica, Universidade 
Estadual de Ponta Grossa, 84030-900, Ponta Grossa, PR, Brazil.}
\address{$^3$Institute for Complex Systems and Mathematical Biology, University
of Aberdeen, AB24 3UE, Aberdeen, SUPA, UK.}
\address{$^4$Departamento de F\'isica, Universidade Federal do Paran\'a, 
81531-990, Curitiba, PR, Brazil.}
\cortext[cor]{Corresponding author: antoniomarcosbatista@gmail.com}

\date{\today}

\begin{abstract}
The brain of mammals are divided into different cortical areas that are 
anatomically connected forming larger networks which perform cognitive tasks. 
The cat cerebral cortex is composed of 65 areas organised into the visual, 
auditory, somatosensory-motor and frontolimbic cognitive regions. We have built
a network of networks, in which networks are connected among themselves 
according to the connections observed in the cat cortical areas aiming to study
how inputs drive the synchronous behaviour in this cat brain-like network. We 
show that without external perturbations it is possible to observe high level 
of bursting synchronisation between neurons within almost all areas, except for
the auditory area. Bursting synchronisation appears between neurons in the 
auditory region when an external perturbation is applied in another cognitive 
area. This is a clear evidence that pattern formation and collective behaviour 
in the brain might be a process mediated by other brain areas under stimulation.
\end{abstract}


\end{frontmatter}


\section{Introduction}
 
The nervous system of mammals is responsible for collecting and processing
information, where the signals are sent by neurons \cite{koch2000}. The 
propagation of neural signals occurs through electrical and chemical synapses, 
as a result of the difference in electric potential between the exterior and 
the interior of a neuron \cite{lent}. Neurons connect to each other forming 
complex layered structures \cite{bullmore2009}. The different cortical layers 
have particular distributions of neuronal cell types, as well as connections 
with other cortical and sub-cortical regions \cite{rockel1980}. The mammalian 
brain is composed of distinct areas, the cerebellar cortex, and non-cortical 
nuclei. The cortex presents fundamental divisions such as the hippocampus 
formation, the olfactory cortex, and associated areas \cite{roland1998}.

In this work we consider the cat cerebral cortex. Scannell and collaborators 
\cite{scannell2,scannell1} have relevant results related to the cortical 
system of the cat. They showed the connection organisation, and reported that 
there are 1139 corticocortical connections among 65 cortical areas. The 
cortical areas are organised into four connectional clusters, corresponding to 
visual, auditory, somatosensory-motor, and frontolimbic areas \cite{scannell3}.

Here we focus on dynamical features such as bursting synchronisation and 
desynchronisation \cite{tonnelier1999}. Bursting synchronisation are thought 
to play relevant roles in information binding in the mammalian brain 
\cite{lestienne01}. However, bursting synchronisation may be associated with 
pathologies like seizures \cite{boucetta2008} or Parkinson's disease 
\cite{schwab13}. For this reason, studies about synchronisation are of great 
interest to neuroscience.

Our purpose in this work is to study the formation of patterns of 
synchronisation and desynchronisation in a neural network model of the cat 
brain, using the matrix of corticocortical connections in the cat 
\cite{scannell1}. The matrix represents the densities of connections in 65 
cortical areas is on undirected weighted adjacency. We describe each cortical
area as a small-world network \cite{watts,watts98,zhou07}. Small-world networks
have been proposed to be an efficient solution for achieving phase 
synchronisation of bursting neurons \cite{batista12}. In addition they have
been found to be linked to different levels of models of the brain.

Small-world networks have been intensively investiga\-ted in computational 
neuroscience \cite{korenkevych2013}. The characterisation can be made on two 
basic levels: a microscopic, neuroanatomic level, and a macroscopic, functional
level. Studies at the former level are limited to those few examples in which 
there is available data on the neuronal connectivity, as the worm 
{\it C. Elegans}, which is considered one of the simplest and most primitive 
organisms that shares essential biological characteristics of the more complex 
species \cite{varshney}.  Moreover, there have been studies of large-scale 
anatomical connection patterns of the human cortex using cortical thickness 
measurements from magnetic resonance imaging \cite{he2007}. The human brain 
anatomical network at this level has an average path length and a clustering 
coefficient with values presented by networks with small-world property. Stam 
and collaborators, in a study of functional brain networks, observed that there
is a loss of small-world network characteristics in patients with Alzheimer's 
disease, in particular with an increase of the average path length with no 
significant changes in the clustering coefficient \cite{stam}. 

At the macroscopic level of description of neural networks, the use of 
non-invasive techniques as electroencephalography, functional magnetic 
resonance imaging and magnetoencephalography provides anatomical and functional 
connectivity patterns between different brain areas \cite{hilgetag,gorka2}. 
This information provides a way to study the brain cortex, considering the 
latter as being divided into anatomic and functional areas, linked by axonal 
fibers. Scannell and coworkers have investigated the anatomical connectivity 
matrix of the visual cortex for the macaque monkey and the cat 
\cite{scannell2,scannell1}. In both cases the values of the average path length
and clustering coefficient are in accordance with expected small-world 
properties \cite{scannell2,scannell1}. 

Each node in our brain model is described by the Rul\-kov model \cite{rulkov}, 
a discrete time system with two dimensions. The low-dimensionality of this map
allow us to study large neural networks of approximately 10,000 neurons.
This model has been extensively tested and it reproduces well the main
features of realistic neural models. It has been used in studies about control 
of bursting synchronisation \cite{batista10}, phase synchronisation in 
clustered networks \cite{batista12}, and suppression of bursting 
synchronisation \cite{lameu12}.

Our main goal is to show that pattern formation and collective behaviour
in the brain might be a process mediated by other brain areas under 
stimulation. We also show that external perturbations induce synchrony 
behaviour in cognitive areas of the cat cerebral cortex. Bursting 
synchronisation appears between neurons in the auditory region when an external
perturbation is applied in another cognitive area. We consider perturbations
that activate neurons in accordance with experimental results in that pulses of
blue light are capable to induce neuronal spikes \cite{boyden1,boyden2}.

This paper is organised as follows: in Section II we introduce the network of
Rulkov neurons and the cat brain matrix. In Section III, we study the 
phase synchronisation of the cognitive areas according to electrical and
chemical synapses. In Section IV, we analyse the effect of an external 
perturbation on the synchronisation. In the last Section, we draw the 
conclusions.


\section{Network of Rulkov neurons}

\begin{figure}[hbt]
\centering
\includegraphics[height=8cm,width=8cm]{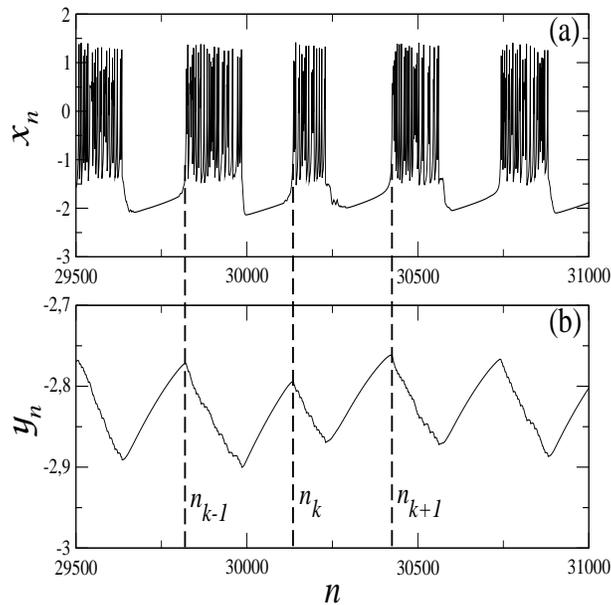}
\caption{Time evolution of the (a) fast and (b) slow variables in the Rulkov 
map.}
\label{rulkov}
\end{figure}

There is a wide range of mathematical models used to describe neuronal activity
\cite{rulkov,hodgkin,fitzHugh,hindmarsh}. In this work we consider the 
phenomenological model proposed by Rulkov  
\begin{eqnarray}
x_{n+1} &=& \frac{\alpha}{1 + x_n^2} + y_n, \\
y_ {n+1} &=& y_n - \sigma(x_n - \rho),
\end{eqnarray}
where $x_n$ and $y_n$ are the fast and slow dynamical variables, respectively.
The parameter $\alpha$ affects the spiking time-scale, and we choose values
in that the time series of $x_n$ presents an irregular sequence of spikes. The 
parameters $\sigma$ and $\rho$ describe the slow time-scale. Figure
\ref{rulkov} shows the time evolution of the fast and slow variables, where 
$n_k$ is used to denote when neuronal bursting starts.

\begin{figure}[hbt]
\centering
\includegraphics[height=10cm,width=10cm]{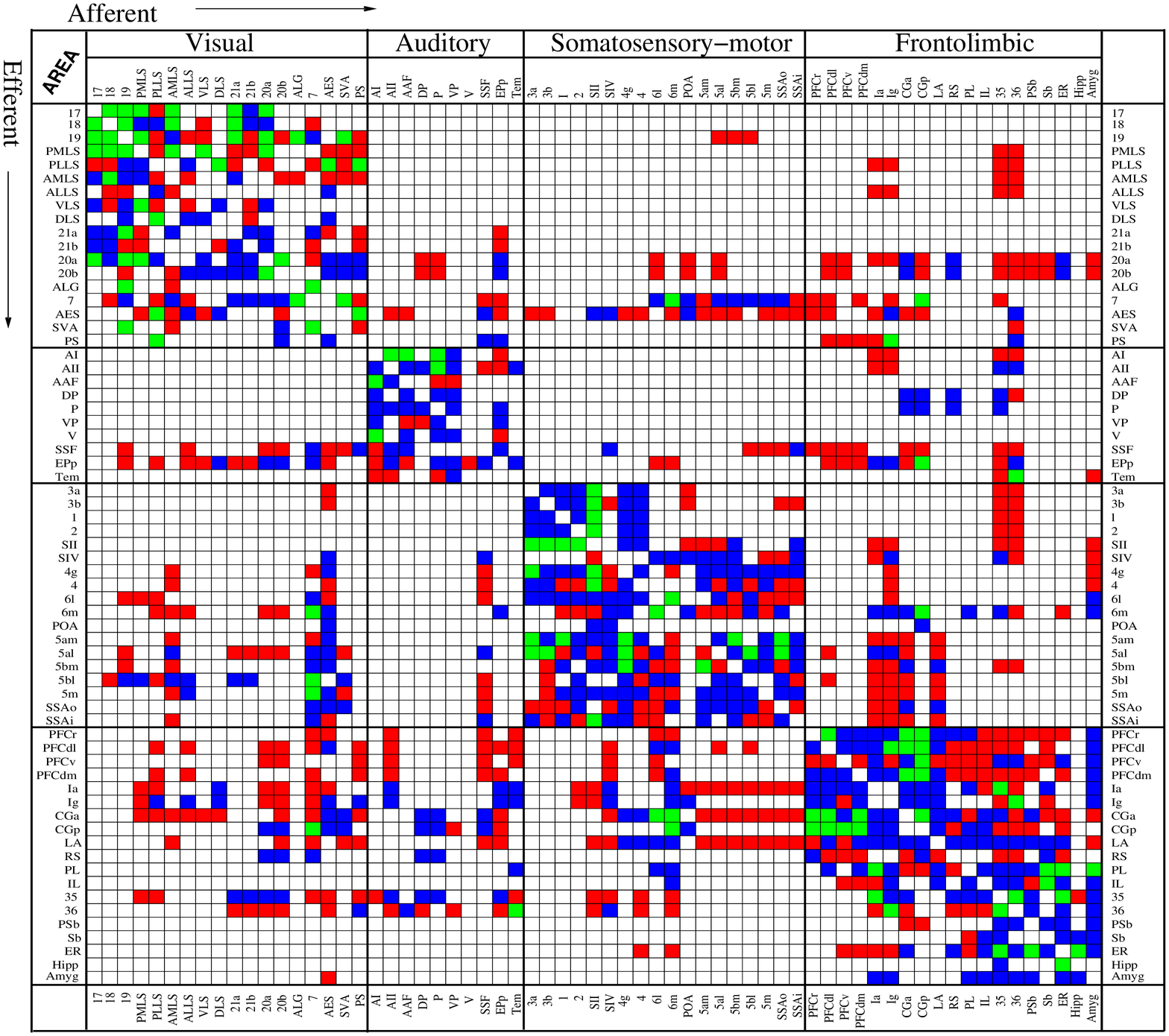}
\caption{(Colour online) Density of connections between cortical areas 
classified as absent of connection (white), sparse (red), intermediate (blue), 
and dense (green).}
\label{cat}
\end{figure}

We build a neuronal network considering the organisation of corticocortical 
connections obtained by Scannell and collaborators  
\cite{scannell1,scannell2,scannell3}. Figure \ref{cat} exhibits the matrix
description of the corticocortical connectivity of the brain in accordance
with Ref. \cite{scannell3}. In this work, the matrix elements have connections 
weighted 0, 1, 2, or 3, where 0 represents the absence of a connection between
two brain areas, 1 represents a sparse connection, 2 represents an intermediate
connection, and 3 is dense connections. In Figure \ref{cat} the colours are 
representing the weights white for no connections, sparse connections in red, 
intermediate connections in blue, and dense connections in green. All areas are
grouped into 4 cognitive regions: visual, auditory, somatosensory-motor, and 
frontolimbic. The visual region has 18 cortical areas, the auditory has 10
areas, the somatosensory-motor has 18 areas, and the frontolimbic has 19 
cortical areas. 

The cortical areas contain a characteristic distribution of neuronal cells and
connections. Supporting evidences that there are small world properties at 
different level models of the brain \cite{sporns,shan} we consider each 
cortical area as a small world network. A small-world has an average distance 
among neurons like a random network, while the degree of clustering is 
comparable to a regular network \cite{watts98}. A small-world network has 
typically an average distance between sites comparable to the value it would 
take on for a random network, while retaining an appreciable degree of 
clustering, as in regular networks. Watts and Strogatz obtained small-world 
networks from an otherwise regular lattice with local connections, to which 
non local connections were added by randomly rewiring a small fraction of the 
local connections \cite{watts98}. An alternative procedure was proposed by 
Newman and Watts, who inserted randomly chosen shortcuts in a regular lattice, 
instead of re-wiring local links into non-local ones \cite{newman99}. We build 
small-world networks according to the procedure proposed by Newman and Watts. 
Each small-world network has 100 neurons and $\%5$ of shortcuts. The 
connections among the small-world networks obey the corticocortical 
connectivity of the cat (Fig. \ref{cat}) so that 2 areas connected with weight 
equal to 1 (red) have 50 randomly connections, areas connected with weight 
equal to 2 (blue) have 100 randomly connections, and areas with weight 3 
(green) have 150 connections. 

The coupling between neurons happens by means of electrical or chemical 
synapses. The local connections between neurons within each small-world 
network are described by electrical synapses. The shortcut non-local 
connections between neurons within each small-world network and the 
connections among cortical areas are described by chemical synapses. Chemical 
synapses may be excitatory or inhibitory. We consider that $75\%$ are 
excitatory and $25\%$ are inhibitory \cite{bannister05}.

The dynamic behaviour of the neuronal network with electrical and chemical
connections is governed by the following equations
\begin{eqnarray}
x_{n+1}^{(i,p)} &=& \frac{\alpha^{(i,p)}}{1+(x_n^{(i,p)})^2}+y_n^{(i,p)} 
+\frac{g_e}{\gamma^{(i,p)}}\sum_{\stackrel{(q,i)\in S}{ p\in P}}E_{(q,p),(i,p)}
(x_n^{(q,p)}-x_n^{(i,p)}) \nonumber \\  
& & -g_c\sum_{\stackrel{(d,i)\in S}{(f,p)\in P}}\left[ A_{(d,f),(i,p)}H(x_n^{(d,f)}-\theta)
(x_n^{(i,p)}-V_s)\right ] +\Lambda_n, \\
y_{n+1}^{(i,p)} & =& y_n^{(i,p)}-\sigma(x_n^{(i,p)}-\rho),
\end{eqnarray}
where $S$ ($i=1,2,...,S$) is the total number of neurons in each small-world,
$P$ ($p=1,2,...,P)$ is the number of cortical areas ($P=65$ according to Fig. 
\ref{cat}), $g_e$ is the electrical coupling strength, $g_c$ is the chemical 
coupling strength, $\alpha^{(i,p)}$ is the non-linearity parameter of the Rulkov
map with values randomly coupled in the interval $[4.1, 4.4]$, $\sigma=0.001$, 
and $\rho=-1.25$. The term $\Lambda_n$ is an external excitatory perturbation
which activates spikes in randomly chosen  neurons. The adjacency matrices 
$E_{(q,p),(i,p)}$ and $A_{(d,f),(i,p)}$ are the electric and chemical connections, 
respectively. They have elements with value equal to 1 when neuron ($q,p$) 
connects electrically with neuron ($i,p$), and neuron ($d,f$) connects 
chemically with neuron ($i,p$). $H(x)$ is the Heaviside step function, where 
$\theta=-1.0$ is the presynaptic threshold for the chemical synapse. When the 
presynaptic neuron voltage is above $\theta$, the postsynaptic neuron receives 
an input. This way, $\theta$ is related to the sharp voltage response of the 
presynaptic terminals. The constant $V_s$ denotes the reversal potential 
associated with the synapse, that is defined by the nature of the postsynaptic 
ionic channels. The synapse will be excitatory if $V_s$ is higher and 
inhibitory if $V_s$ is lower than a specific range \cite{ibarz11}. Regarding 
the network of Rulkov neurons, for excitatory synapses $V_s=1.0$ and for 
inhibitory $V_s=-2.0$.


\section{Burst synchronisation without external perturbation}

Bursts of spikes and oscillatory patterns of neuronal activity have been 
observed in the central nervous system, ranging from slow to fast oscillations 
\cite{buzsaki06}. In fact, synchronised bursting has been verified in EEG 
recording of electrical brain activity. A type of synchronisation is the burst
phase synchronisation \cite{batista07}. 

Burst phase synchronisation is studied through the definition of a phase for 
neuronal bursting, defined by the slow variable. We consider that a burst 
begins when the slow variable $y_n$, shown in Fig. \ref{rulkov}(b), has a local
maximum happening in a time $n_k$. The duration of the burst, $n_{k+1}-n_k$, 
depends on the variable $x_n$ and fluctuates in an irregular behaviour as long 
as $x_n$ undergoes irregular evolution. Then, we define a phase describing the 
time evolution within each burst, varying from 0 to $2\pi$ as $n$ evolves from 
$n_k$ to $n_{k+1}$,
\begin{equation}
\phi_n=2\pi k+2\pi \frac{n-n_k}{n_{k+1}-n_k},
\end{equation}
where $k$ is an integer.

We use the Kuramoto's order parameter as a diagnostic of the burst phase 
synchronisation, that is defined as
\begin{equation}\label{porder}
z_n^{(l)}=R_n^{(l)}\exp ({\rm i}\Phi_n^{(l)})\equiv 
           \frac{1}{N_{l}}\sum_{j\in I_{l}}\exp ({\rm i}\phi_n^{(j,I_l)}),
\end{equation}
where $R_n$ is the amplitude and $\Phi_n$ is the angle of a centroid phase 
vector for an one-dimensional network with periodic boundary conditions.
$I_{l}$ represents one of the four cognitive areas, $l=1$ for visual,
$l=2$ for auditory, $l=3$ for somatosensory-motor, and $l=4$ for
frontolimbic. $N_{l}$ is the number of neurons of each area. 
$\phi_n^{j,I_l}$ represents the phase of the neurons $j$ belonging to the c
cortical area $I_l$. If the bursting phases $\phi_n^{(j,I_l)}$ are uncorrelated, 
the summation in Eq. (\ref{porder}) is small and $R_n^{(l)}\ll 1$. Whereas 
$R_n^{(l)}=1$ when the cortical network area is in a completely burst phase 
synchronised state. 

We are interested in studying the role of the electrical and chemical coupling 
strength at the level of bursting synchronisation. For this reason, we use the 
time averaged order parameter magnitude, given by
\begin{equation}
{\bar R}^{(l)}= \frac{1}{T}\sum_{n=1}^T R_n^{(l)},
\end{equation}
where $T$ is the time interval, as a measure of synchronisation in the network.
If the bursting dynamics in the area $I_l$ is globally synchronised, we obtain 
$\overline{R}^{(l)} \approx 1$. 

\begin{figure}[hbt]
\centering
\includegraphics[height=9cm,width=11cm]{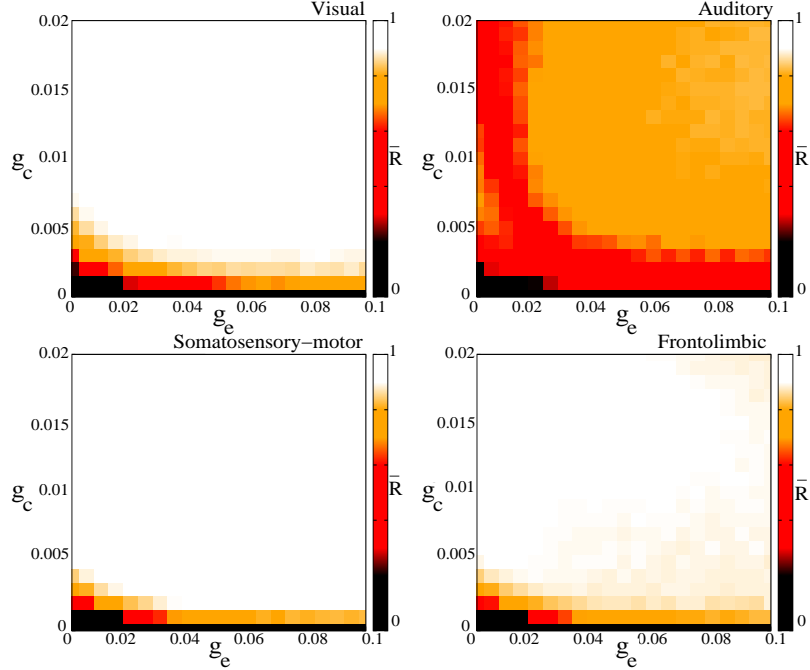}
\caption{(Colour online) Average order parameter of the cat's cognitive brain 
regions. We consider 50000 iterations where 20000 were transient.}
\label{areas}
\end{figure}

Figure \ref{areas} shows the time averaged order parameter in colour scale as a 
function of the electrical and chemical coupling strength. Due to the fact
that the neurons are not identical the neuronal network does not present
a completely phase synchronised state (${\bar R}^{(l)}\approx 1$). However, 
the network exhibits strong synchronisation for ${\bar R}^{(l)}>0.9$, 
corresponding to the white region. All the cognitive areas, except for the 
auditory, present strong synchronisation (white region).

The auditory region does not present high levels of bursting synchronisation.
This effect is due to the complex network topology, where the synchronisation
patterns are controlled by input intensities among neurons, and also among the
corticocortical areas \cite{zhou07}. To show that we define the mean field of 
the auditory area
\begin{equation}
M_n^{(2)} = \frac{1}{N_2}\sum_{(i,p)\in I_2}x_n^{(i,p)}, 
\end{equation}
and the mean field of the $N_{\rm out}$ inputs coming from other areas
\begin{equation}
C_n^{(l)} = \frac{\Im_n^{(l)}}{N_{\rm out}},
\end{equation}
where 
\begin{equation}
\Im_n^{(l)} = -g_c \sum_{\stackrel{(i,p)\in I_2}{(d,f)\in I_1,I_2,I_3}} 
\left[ A_{(d,f),(i,p)} H(x_n^{(d,f)}-\theta)(x_n^{(i,p)}-V_s)\right],
\end{equation}
where $N_{\rm out}$ represents the number of non null elements in the matrix
$A_{(d,f),(i,p)}$ for $(i,p)\in I_2$ and $(d,f)\in I_1,I_2,I_3$. We assume that 
each neuron $(i,p)$ belonging to a set $I_2$ in the auditory area is connected 
to a neuron $(d,f)$ in a set $I_{l}$ belonging to another cognitive area. Due 
to spatial distance between areas \cite{beul14} we consider that the inputs 
from other areas are through chemical synapses \cite{pereda14}.

\begin{figure}[hbt]
\centering
\includegraphics[height=8cm,width=10cm]{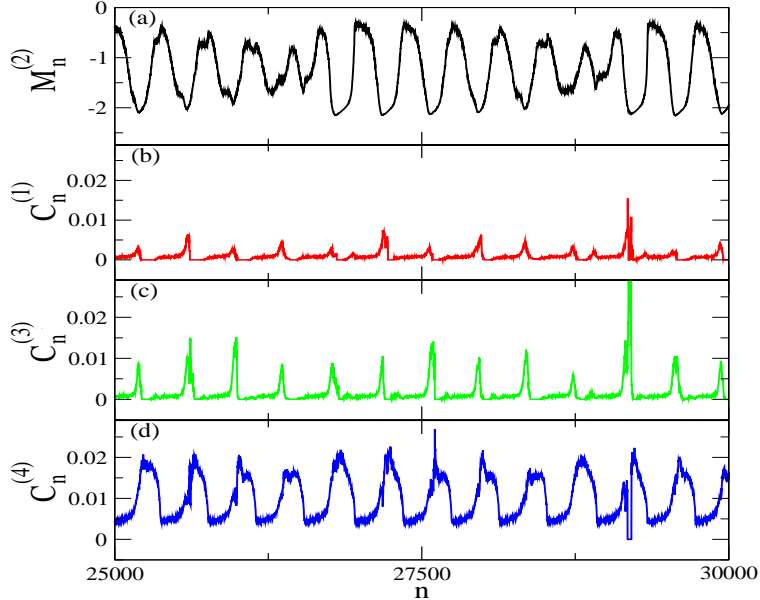}
\caption{(Colour online) (a) Dynamics of the auditory fast variable mean field.
Evolution of the external mean field on auditory area from (b) visual, (c) 
somatosensory-motor, and (d) frontolimbic areas. We consider $g_e=0.05$ and 
$g_c=0.015$.}
\label{fig4}
\end{figure}

\begin{figure}[hbt]
\centering
\includegraphics[height=8cm,width=10cm]{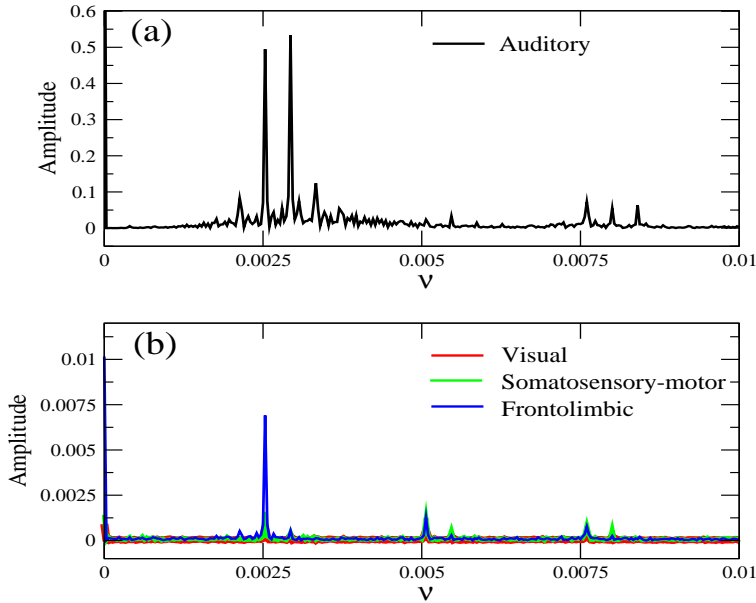}
\caption{(Colour online) Frequencies of the (a) auditory fast variable mean 
field, and (b) external mean field on auditory area, considering $g_e=0.05$ and
$g_c=0.015$.}
\label{fig5}
\end{figure}

Aiming to demonstrate that the absence of high levels of synchrony in the 
auditory area is an effect of the network, we compare the mean fields 
$M_n^{(2)}$ with $C_n^{(1)}$, $C_n^{(3)}$, and $C_n^{(4)}$. Figure \ref{fig4}(a) 
exhibits the time evolution of the mean field $M_n^{(2)}$ for $N_2=1000$. The 
irregular time evolution, that it is associated with a non synchronised 
behaviour, is influenced by the stimuli from the visual (Fig. \ref{fig4}b), the
somatosensory-motor (Fig. \ref{fig4}c), and the frontolimbic (Fig. \ref{fig4}d)
areas. This figure shows that the collective behaviour of the auditory area is
correlated with the input from the frontolimbic. If the auditory area is 
isolated from other areas, by making $A_{(d,f)\in I_1,I_3,I_4;(i,p)}=0$ it will 
present global bursting synchronisation with a frequency around 0.0027. The 
frequency is obtained from bursts. Connecting the auditory area with the rest 
causes a suppression of bursting synchronisation, as well as the appearance of 
one more frequency with value equal to 0.0025 (Fig. \ref{fig5}a). We can see 
through Figure \ref{fig5}(b) that the new frequency appears by a resonant 
effect caused by the oscillations of the stimulus from others areas. The visual
and somato-motor areas present the frequency equal to 0.0025 with magnitude 
equal to 0.0005 and 0.001, respectively. The frontolimbic area also exhibits 
this value of frequency, but the magnitude is larger than visual and 
somato-motor areas. Therefore, the low levels of synchrony in the auditory area
is mainly due to the new frequency of the input from frontolimbic area.

\section{External perturbation inducing synchronisation}

It has been experimentally found that the incidence of light with a suitably 
chosen frequency on a group of neurons is able to alter their spiking activity.
This phenomenon was verified by Boyden and collaborators 
\cite{boyden1,boyden2}, by means of blue light they activated spikes in neurons
genetically modified, and also suppressed spikes through yellow light. With 
this in mind, we consider an external perturbation ($\Lambda_n$) acting on 100 
neurons randomly chosen in the same cognitive area. This perturbation, aiming 
to simulate blue light optical stimulus, has excitatory effect on the perturbed
neurons making them spike independent of their previous state.

\begin{figure}[hbt]
\centering
\includegraphics[height=9cm,width=11cm]{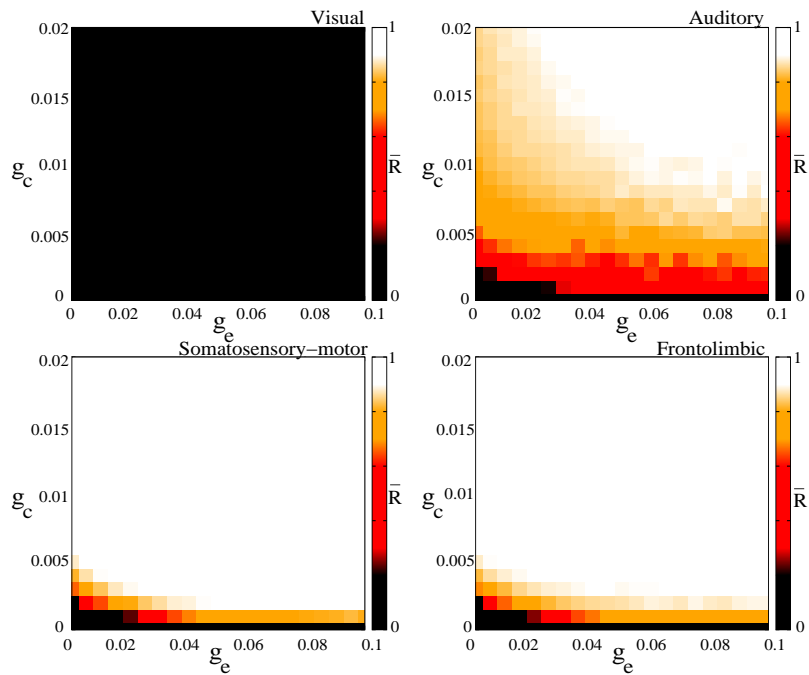}
\caption{(Colour online) Order parameter of the cat's cognitive brain regions 
with an external perturbation on the visual area. We consider 50000 
iterations, and 20000 transient iterations.}
\label{fig6}
\end{figure}

We apply an external perturbation on the visual area to verify the 
synchronisation behaviour. Figure \ref{fig6} shows the effects of this 
perturbation in the cognitive areas. Synchronisation is fully suppressed in the 
visual area. The somatosensory-motor and frontolimbic areas do not present 
significant alterations due to the perturbation in the visual area. This 
behaviour is due to the small number of connections between these areas and
the visual area. On the other hand, the auditory area exhibits a change in its 
synchronous behaviour.  Without an external perturbation the auditory area does 
not present high levels of synchronisation, shown in Figure \ref{areas}, but 
with the perturbation in the visual area it is possible to observe strong
synchronisation domains in the auditory area.

\begin{figure}[hbt]
\centering
\includegraphics[height=8cm,width=10cm]{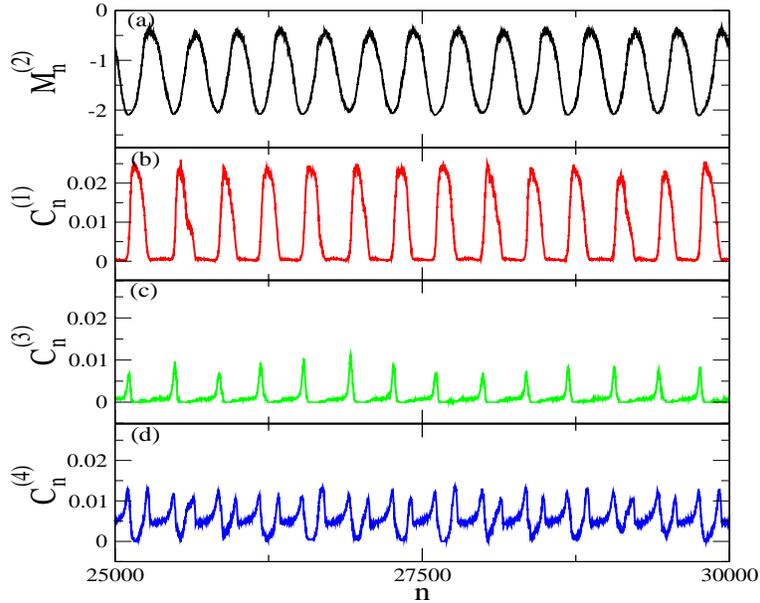}
\caption{(Colour online) (a) Dynamics of the auditory fast variable mean field.
Evolution of the external inputs mean field on the auditory area from (b) 
visual, (c) somatosensory-motor and (d) frontolimbic areas with an external 
perturbation on the visual area, considering $g_e=0.05$ and $g_c=0.015$.}
\label{fig7}
\end{figure}

\begin{figure}[hbt]
\centering
\includegraphics[height=8cm,width=10cm]{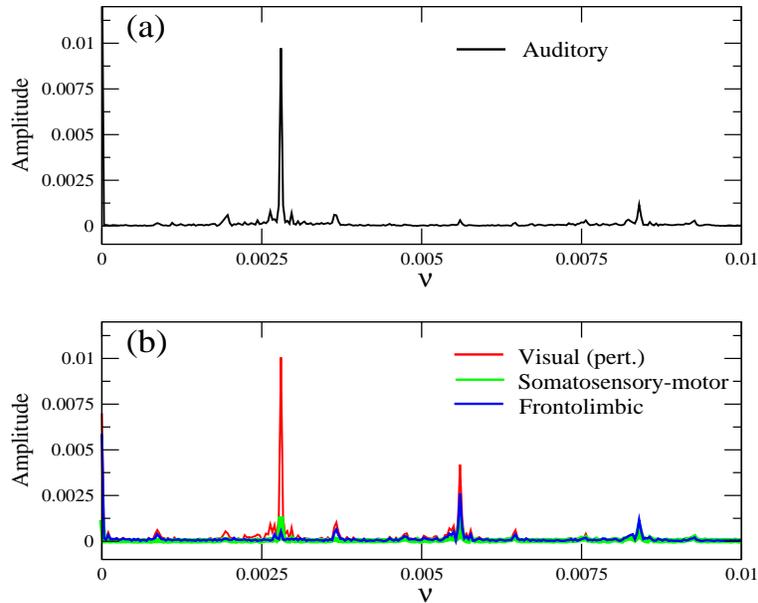}
\caption{(Colour online) Frequencies of the (a) dynamics of the auditory fast 
variable mean field and (b) inputs mean field on auditory area for the system 
with external perturbation on visual area.}
\label{fig8}
\end{figure}

In order to understand the perturbation effect in the auditory area we calculate
the time evolution of the mean field $M_n^{(2)}$, and the mean field from 
inputs coming from others areas $C_n^{(1)}$, $C_n^{(3)}$, and $C_n^{(4)}$
into the area $I_2$. Figure \ref{fig7}(a) shows that the mean field of the 
auditory area has a regular behaviour, whereas with no external perturbation 
the behaviour is irregular. Comparing the results of Figure \ref{fig7} for the 
external inputs of the mean field on auditory from visual (b), 
somatosensory-motor (c), and frontolimbic (d) with the results of Figure 
\ref{fig4} we can see that the inputs from the visual area becomes 
frequency-locked with the mean field $M_n^{(2)}$ of the auditory area. 
$M_n^{(2)}$ and $C_n^{(1)}$ present both the same frequency, approximately 
$0.0028$, as it can be seen in Fig. \ref{fig8}. 

Perturbations applied in auditory area do not alter significantly the 
synchronous behaviour of the other areas. Moreover, we have verified which 
perturbations in the somatosensory-motor and frontolimbic areas produce the 
same effect on the network such as when they are applied in the visual area. 


\section{Conclusion}

In this paper we studied burst phase synchronisation in a neuronal network with
a topology according to the corticocortical connections of the cat cerebral 
cortex. When no perturbation is applied we verified that only the auditory area
does not present synchronisation in the parameter space. This happens because 
of the large influence of the other areas. The ratio between inter connections 
and intra connections is less for the auditory area than other areas. As the 
consequence the frontolimbic area induces a frequency in the auditory area 
which helps to suppress burst phase synchronisation.

We verified the suppression of burst phase synchronisation when a stimulus is 
applied on a cognitive area. However, the suppression in a specific area can 
affect another area. Our results have showed that an external perturbation 
applied into the visual area suppresses synchronisation in the visual area, but
surprisingly it induces synchronisation in the auditory area. The same happens
for perturbation applied in the somatosensory-motor and frontolimbic which 
induces synchronisation in the auditory area. 


\section*{Acknowledgements}
This study was possible by partial financial support from the following 
Brazilian government agencies: CNPq, CAPES, Science Without Borders Program.
Murilo S. Baptista also acknowledges EPSRC-EP/I032606/1.

\end{document}